\documentstyle{mn}
\title{A search for hidden white dwarfs in the {\RO} EUV survey$^\dag$}

\author[M.\,R. Burleigh et al. ]
{M.\,R. Burleigh$^1$$^*$, M.\,A. Barstow$^1$$^*$ and T.\,A. Fleming$^2$ \\
$^1$ Department of Physics and Astronomy, University 
of Leicester, University Rd., Leicester, LE1 7RH \\
$^2$ Steward Observatory, University of Arizona, Tucson, Arizona 85721,
USA \\
$\dag$ Based on observations made with the {\RO} observatory and the
International Ultraviolet Explorer ({\IUE}) satellite \\
$^*$Guest Observers with the International Ultraviolet Explorer ({\IUE})
satellite \\
}
\date{January 11th 1997}
\def\RO{\it ROSAT\rm }
\def\euve{\it EUVE\rm }
\def\IUE{\it IUE\rm }
\def\TD-1{\it TD-1\rm }
\def\tkev{\thinspace{ke\kern-.15em V}}
\def\tev{\thinspace{e\kern-.15em V}}

\begin{document}
\maketitle
\begin{abstract}

The {\RO} WFC survey has provided us with evidence for the existence of a
previously unidentified sample of hot white dwarfs (WD) in non-interacting 
binary systems, through the detection of
EUV and soft X-ray emission. These stars are hidden at optical
wavelengths due to their close proximity to much more luminous main
sequence (MS) companions (spectral type K or earlier). However, for companions
of spectral type $\sim$A5 or later the white dwarfs are
easily visible at far-UV wavelengths, and can be identified in spectra
taken by {\IUE}. Eleven white dwarf binary systems have previously been found 
in this way from 
{\RO}, {\euve} and {\IUE} observations (e.g. Barstow et al. 1994). 
In this paper we
report the discovery of three more such systems through our programmes in
recent episodes of {\IUE}. The new binaries are HD2133, RE
J0357$+$283 (whose existence was  
predicted by Jeffries, Burleigh and Robb 1996), and 
BD$+$27$^\circ$1888. In addition, we have independently identified a
fourth new WD$+$MS binary, RE J1027$+$322, which has also been reported
in the literature by Genova et al. (1995), bringing the total number of
such systems discovered as a result of the EUV surveys to fifteen.  
We also discuss here six stars which were observed as part of the
programme, but where no white dwarf companion was found. Four of these
are coronally active. Finally, we present an analysis of the WD$+$K0IV
binary HD18131 (Vennes et al. 1995), which includes the {\RO} PSPC X-ray
data.

\end{abstract}


\section {Introduction}

The extreme ultraviolet (EUV) surveys of the {\RO} Wide Field Camera
(WFC, Pounds et al. 1993, Pye et al. 1995) and the Extreme Ultraviolet
Explorer ({\euve}, Bowyer et al. 1994, 1996) have found a substantial
number of hot white dwarfs ($\approx
120$). The majority of these are single isolated stars, but about 30
are now known to lie in binary systems. Some have been identified
because they are in interacting Cataclysmic Variable systems. Others have
been found,  from optical observations, to be 
in pairs with faint red dwarf companions (e.g. RE J1629$+$780, Cooke et al.
1992). A few are in wide, resolved  binaries (e.g. HD74389B, Liebert,
Bergeron and Saffer 
1990). However, unresolved, non-interacting pairs with companions earlier
than M-type have remained very difficult to identify.

In optical surveys, white dwarfs  have been identified on the basis of
colour information (e.g. Green, Schmidt and Liebert 1986), and there
is thus a selection effect against the detection of those in unresolved
binaries.
Any companion star of type K or earlier will completely dominate the
optical spectrum of the white dwarf (see Figure 15). 
For example, Sirius B would be
optically undetectable were it not for the close proximity of the system
to Earth (2.64pc), allowing it to be resolved from Sirius A.

Several unresolved systems have been discovered
serendipitously. The white dwarf in the well-studied system V471 Tauri
was found as a result of an eclipse by its K2V companion (Nelson \&
Young 1970). A number of others have been  found by chance in
ultraviolet (UV) spectra taken by the 
International Ultraviolet Explorer ({\IUE}).    
For example, the white dwarf companion to HD27483 (B\"ohm-Vitense 1993)
was discovered as part of an {\IUE} study of Hyades F stars by
B\"ohm-Vitense (1995). Others detected in a similar fashion include    
 $\zeta$ Cap (B\"ohm-Vitense 1980),  56 Peg (Schindler et al. 1982)  
and 4 $o$$^1$ Ori (Johnson \& Ake 1986). In 1989, Shipman and
Geczi  conducted a systematic survey of the then existing {\IUE} 
archive for white dwarf companions to G, K and M stars, but found no
further examples.

Recently Landsman, Simon and Bergeron (1996) have detected, using {\IUE},   
two white dwarf companions to F stars showing an excess of ultraviolet 
radiation  
in the {\TD-1} sky survey (Thompson et al. 1978). The white dwarf in the 56
Persei system (F4V) is, at T$=$16,400K, too cool to be seen with the WFC.
The white dwarf companion to HR3643 is hot (29,000$<$T$<$36,000), but
it is not detected by either {\euve} or {\RO}. This is probably because of a
high interstellar hydrogen column ($\sim$2$\times$10$^{20}$cm$^{-2}$) to this
star. 

{\RO} and {\euve} have now provided us with evidence for the existence of
many more of these hidden hot white dwarfs through the detection of
EUV and soft X-ray emission. 
Follow-up observations in the UV with {\IUE} have enabled detection 
of unresolved white dwarfs in systems with a companion later than
spectral type $\sim$A5. The initial discovery of companions to $\beta$ Crateris
(A2IV, Fleming et al. 1991) and HD33959C (KW Aur C, F4V, 
Hodgkin et al. 1993) was followed
by seven others, all discussed in Barstow et al. (1994):
BD$+$$08^\circ$102 (K1-3V, also Kellett et al. 1995), HR1608 (K0IV, also
Landsman, Simon, Bergeron 1993, hereafter LSB), HR8210 (IK Peg, A8m,
also LSB, Wonnacott et al. 1993, Barstow, Holberg and Koester 1994), 
HD15638 (F3-6V, also LSB, Barstow, Holberg and Koester 1994), 
HD217411 (G5), HD223816 (F5-G0), and RE J1925$-$566 (G2-8).
Since then Vennes et al. (1995) have reported a DA companion to the active K0IV
star HD18131 through its detection as an {\euve}/WFC source, and
Christian et al. (1996) have discovered yet another DA$+$G star binary,
MS 0354.6$-$3650, which is also an {\euve} source (although it does not
appear in the {\RO} WFC catalogues). This work has
clearly demonstrated that there is an as yet unexplored population of
white dwarfs with companions earlier than type M.

Most of these white dwarfs are all relatively 
bright EUV sources (with the exceptions of BD$+$$08^\circ$102 and MS
0354.6$-$3650), with
distinctive spectral signatures. Typically, the S2/S1 count rate ratios
are $>$2. For
example HD33959C has a count rate of 2.628s$^{-1}$ in the WFC S2 filter,
and 1.172s$^{-1}$ in S1. Comparisons with known white dwarfs detected by
the WFC (e.g. GD659 2.193s$^{-1}$ in S2, 0.722s$^{-1}$ in S1) 
left little doubt that these sources were indeed white dwarfs. 
The observations with {\IUE} merely confirmed the identifications. 

About 120 of the 383 sources in the original Bright Source Catalogue 
(BSC, Pounds et al. 
1993) have been associated with hot white dwarfs, and a larger number
($\sim 180$) with active late type stars.
However, from a cursory glance at the original BSC, 
it is clear that some of these late type stars were not previously known
to be chromospherically active. 
Indeed, follow-up optical studies (e.g. Mason et al. 1995, Mullis and
Bopp 1994) failed to find any evidence of activity at all in a few
cases.  It is also obvious from the WFC catalogues that there
are many faint hot white dwarfs with low count rates, and with EUV
colours that are not particularly distinctive. 
These stars may lie in relatively high column 
directions that affect the S2/S1 ratio and, from the count rates alone, 
it is difficult to immediately distinguish them from an active star. 
Furthermore, some of 
the primaries in the binaries already discovered are also active (e.g.
HD18131, BD$+$$08^\circ$102 and RE J1925$-$566). Determining which EUV source 
is due to an active star and which is due to a white dwarf is clearly, in
some cases, not an easy task. Given all these factors
we continued to use {\IUE} to search  
for further, less obvious white dwarfs hidden in non-interacting binaries.
Since we began our {\IUE} programmes 
the 2nd {\RO} WFC catalogue (Pye et al.,
1995) has been published. This utilises improved source detection methods
and contains 120 new sources as well as updating the BSC identifications.
There is thus a pool of new, fainter and less obvious candidates 
requiring observation.    

In this paper we report the results of our latest searches with {\IUE}. 
We have identified four more hot white dwarfs in
non-interacting binary systems with main sequence stars earlier than type M. 
They are HD2133 (RE J0024$-$741), 
RE J0357$+$283, BD$+$27$^\circ$1888 (RE J1024$+$262) and RE J1027$+$322.
Jeffries, Burleigh and Robb (1996) had previously
predicted  a hidden hot white dwarf as the most likely
source of the EUV radiation in RE J0357$+$283. 
RE J1027$+$322 has also been independently reported by Genova et al. (1995); 
we present further observations and analysis of this system. 
This brings the total number of these WD$+$MS binaries 
discovered as a result of EUV surveys to fifteen.

We also provide in this paper an analysis of the white
dwarf companion to HD18131, discovered and reported by Vennes et
al. (1995). We include X-ray data from the {\RO} Position Sensitive
Proportional Counter (PSPC), which were not available to Vennes et al.
Those authors had observed Mg II emission lines in {\IUE} LWP spectra
of the K0IV primary, suggesting that it is active.  
We detect X-ray radiation from this system in the PSPC upper band  
which could not have come from the white dwarf, confirming that the
K0IV star is indeed active. 

In addition, we report here observations of six EUV sources where no white dwarf
was found. Four of these `non-detections' (BD$-$00$^\circ$1462, CD$-$44
1025, HR1249 and HR2468) are active stars, and have been
studied by others both in the UV and the optical. 
We summarise the properties and observations of these
stars, and compare them with the binaries. We note that their UV spectra
show evidence of activity (e.g. CIV and CII emission features) and
compare the level of activity with other known active stars observed with
the WFC. In two cases (AG$+$68 14, HD166435) there is no evidence of
activity in the {\IUE} spectra, and we find no reports in the literature of
these being active stars. We therefore conclude that the counterparts to
the EUV sources (2RE J0014$+$691 and  RE J1809$+$295 respectively)
may lie elsewhere in the fields of these stars and further observations 
are necessary. 

The vast majority of the $>$1500 known white dwarfs are isolated stars.
Theory suggests, however, that over half of all stars should be in binary
or multiple systems. 
The presence of this new population of white dwarfs in binaries 
has profound implications, therefore, for our knowledge of the intrinsic  
white dwarf luminosity function,
formation rate and space density, as determined by e.g. Fleming, Liebert
and Green (1986). Observations of white dwarfs in binaries also
allows us to place constraints on binary evolution models (e.g. de Kool
and Ritter 1993).

\section {Observations}   

\subsection {Detection of the sources in the {\RO} and {\euve} surveys}

The {\RO} WFC EUV and X-ray all-sky surveys were conducted between
July 1990 and January 1991; the mission and instruments are described
elsewhere (Tr\"umper  1992, Sims et al. 
1990). The sources discussed in this paper are all listed in the original
WFC Bright Source Catalogue (Pounds et al. 1993) and in the 2RE
Catalogue (Pye et al. 1995). The revised catalogue contains 479 sources,
as compared to 383 in the original survey. The complete survey database
was reprocessed with improved methods for source detection, better
background screening, etc., to give the 2RE Catalogue. The 2RE
 count rates are quoted in this paper. These are equivalent, on axis
`at-launch' values. The PSPC count rates for all the
white dwarfs detected in the X-ray survey, including the binaries discussed here,
 are given in Fleming et al. (1996). The PSPC count rates for the active
stars (i.e. the non-detections of white dwarf binaries) discussed in 
section 5.2 were obtained via the World Wide Web from the on-line {\RO} All-Sky
Survey Bright Source Catalogue, maintained by the Max Planck Institute in 
Germany (Voges et al., 1997). 

Some of the sources were also detected by {\euve} 
in its all-sky survey (Bowyer et al. 1994, 1996). 
In Table 1 we list
all the {\RO} and {\euve} detections and count rates for our sources. The
quoted {\euve} count rates are from the Second {\euve} Source Catalog (Bowyer
et al. 1996), which includes an improved all-sky detection method
offering better detection sensitivity and reliability.  

\subsection {Selection and Identification of White Dwarfs in Unresolved
Binaries}

The hot white dwarfs detected in the {\RO} EUV and X-ray all-sky surveys
typically have very soft spectra compared to normal stars,
particularly when the hydrogen column is low. The ratio of the WFC survey S2
to S1 count rates can sometimes exceed a factor 2. Additionally, no
photons will be detected from a white dwarf above the 0.28 keV carbon
K$_\alpha$ edge of the {\RO} PSPC. All other
objects generally have spectra extending to higher energies. 

White dwarfs with red dwarf companions are easily identified from their
composite optical spectra, but for binaries with companions of 
spectral type K or earlier the white dwarf cannot be discerned from an optical
observation (see Figure 15). 
However, it is possible to distinguish between the two stars
with a far-UV spectrum with the {\IUE} short wavelength (SWP) camera. Thus,
initially, the stars we chose to observe with {\IUE} were selected by the
similarity of their EUV colours and luminosities to known isolated 
white dwarfs in the WFC survey, after the
elimination of field white dwarfs in chance alignments with late-type
stars, and known EUV active stars. Nine non-interacting binaries were
discovered in this way and discussed by Barstow et al. (1994).

However, only a minority of stars, where the interstellar absorption is
relatively low, have these characteristics. Typically the first of these
systems to be identified were among the brightest EUV sources, but most 
isolated white dwarfs are detetcted at low signal-to-noise ratio (SNR),  
and are  
indistinguishable from coronal sources from EUV data alone. Clearly,
these binaries represent only part of the possible population. 

Only the most active late-type stars are detected as EUV sources. Any that
are weakly active must be very nearby to be seen. Thus RE J1027$+$323 and 
BD$+27$$^\circ1888$ were included on our original {\IUE} target list precisely
because they were not known to be active. During the EUV source optical
identification programme (Mason et al.  1995) observations were made of
possible non-degenerate counterparts to see whether they were sufficiently
active, as indicated by the strength of CaII H and K and H$\alpha$ emission
cores, to be the EUV sources. 
HD2133 was listed in the WFC bright source catalogue as active,
but Mason et al. (1995) found no evidence of activity and thus the most likely
explanation for the EUV flux was a hidden white dwarf companion.
Similarly Mullis and Bopp (1994) found that HD166435 exhibits no
measurable emission in the core of H$\alpha$, and could not conclude 
that it was chromospherically active. 

RE J0357$+$283 was observed by Jeffries, Burleigh and Robb (1996) with the 
LWP camera on {\IUE} (LWP26441) in an attempt to
detect chromospheric MgII H and K emission. An excess of flux
blueward of 2800{\AA} was discovered, strongly suggesting the presence of a
hot white dwarf. No MgII H and K emission was seen, and
an SWP spectrum was needed to conclusively establish the presence of a white
dwarf.

\subsection {UV spectroscopy}

A log of all the UV observations of these objects is given in Table 2.
RE J1027$+$323 and BD$+27$$^\circ1888$ were first observed with {\IUE} in
January 1994 as part of our original search for these systems. Repeat
observations of both stars were later made to improve the quality of
the data through co-addition of the spectra. HD2133 was 
observed in August 1995, and again in October 1995. 
RE J0357$+$283 was added to the programme and observed in August 1995 to
confirm the presence of the white dwarf predicted by Jeffries, Burleigh
and Robb (1996).  
In each case (Figures 1-5) a white
dwarf is clearly identified from the blueward rising flux in the SWP
spectrum. The six non-detections were all observed in this same period. 

\subsection {Optical spectroscopy}

Optical spectroscopy of RE J1027$+$323, HD18131 and RE J0357$+$283 
is reported by Genova et al. (1995), Vennes et al. (1995) and 
Jeffries, Burleigh and Robb (1996) respectively, and we quote their
results here. BD$+27$$^\circ1888$ was observed with the Steward Observatory
2.3m telescope on Kitt Peak, Arizona (Figure 15). 
We have no optical data for HD2133. 
Table 3 lists the parameters of the main sequence stars in the white
dwarf binaries. 

In Table 4 we list some of the properties of the
stars observed where no white dwarf was detected. We have not observed
any of these stars optically ourselves, but summarize results from other
authors.

\section {Data Reduction}

The majority of the far-UV spectra were obtained at {\IUE} Vilspa.
The standard IUESIPS processing was used, with the addition of the following:
a white dwarf based absolute calibration including an effective area 
correction (Finley, 1993), and a correction for the degradation of the 
detector with time (Bohlin and Grillmair, 1988). Garhart (1992) used more
recent {\IUE} images to confirm that the degradation of the SWP camera
sensitivity remained linear with time. In 1995 first
Goddard, and then from October 1995 Vilspa, began using the NEWSIPS
calibration which incorporates these corrections. The NEWSIPS
calibrated data is indicated in Table 2. 
Multiple exposures were obtained for HD2133, RE J1027$+$322, 
BD$+27$$^\circ$1888 and the `non-detection' CD$-$44 1025, and  
co-added before analysis by weighting  each data set according 
to the relative exposure times.

\section {Analysis}

\subsection {White Dwarf Binaries - UV data}

In general, our analysis method follows that of Barstow et al. (1994). The
{\IUE} data can be used to estimate the temperature and gravity of the white
dwarf by fitting the observed Lyman $\alpha$ profile and the
uncontaminated UV continuum to synthetic white dwarf spectra. Obviously
in binary systems like these, there is no possibility of fitting the
Balmer line profiles in the optical region for measurement of T and log
g. Unfortunately, the Lyman $\alpha$ information alone can give somewhat
ambiguous results. For example, fitting the {\IUE} SWP spectrum of the
almost pure H hot 
white dwarf HZ43 yields T$=$57,500K for log g$=$8.5, whereas fitting the
Balmer lines gives a lower T and log g of 49,000$\pm$2000K and
7.7$\pm$0.2 (Napiwotzki et al. 1993).  

We compare the observed UV data with synthetic spectra for grids of fully 
line blanketed LTE homogeneously mixed model
atmospheres (Table 5), spanning  a
temperature range from 20,000K to 100,000K and log g$=$7.0 to 9.0, supplied
by Detlev Koester (e.g. Koester 1991). The models are assumed to be pure
H.

The spectral fitting is conducted with the programme XSPEC
(Schafer et al. 1991), which calculates a chi-squared statistic for the
fit between the data and the model, 
and which is then minimised by incremental steps
in the free parameters. We fit the Lyman $\alpha$ profile and the region
of the UV continuum which is uncontaminated by the primary star. Only in 
BD$+27$$^\circ1888$ is there significant contamination from the companion
in the SWP spectrum (Figure 4). Comparison with stars of similar spectral
type (A8V-F2V) from the {\IUE} archive shows that the flux is
significantly greater 
that of the white dwarf at $\approx$ 1800{\AA} but is effectively zero
at 1500{\AA}. Thus for BD$+27$$^\circ1888$ we fit the UV continuum up to
1500{\AA}.   
We suspected there might be a small amount of contamination in
the SWP from the primary in HD2133 (Figure 1). From an {\IUE} LWP spectrum
(LWP31757) we estimate the spectral type of this star to be F7V-F8V. 
Comparison with example spectra from the {\IUE} archive shows that
there will be a contribution from a companion of this type of a few $\%$ 
of the white dwarf flux at 1800{\AA}, but none at 1600{\AA}. We therefore fit 
the UV continuum up to 1600{\AA}.  

In the IUESIPS data sets, the Lyman $\alpha$ profile is 
sometimes significantly contaminated
by geocoronal radiation, and only the outer wings can be used. In the
NEWSIPS 
data, this geocoronal line is removed during the calibration process. 
However, we find significant differences in the fits to
BD$+$27$^\circ$1888 
from the NEWSIPS and the IUESIPS data. This problem is discussed later in
section 5.1.4. 

There is no
need to take into account any interstellar component in the fits to the
Lyman $\alpha$ profile, because for low
columns this will be negligible, and for columns of a few $\times$10$^{19}$
the contribution is still lower than the typical uncertainties in the
observed fluxes, between $\approx$ 5$-$10$\%$. For columns greater than a
few $\times$ 10$^{19}$ the white dwarf is unlikely to be detected in EUV surveys. 

The errors on the flux values are derived using the
general scatter from a smooth curve passing through
the SWP spectrum, since our own calibrations, unlike the NEWSIPS
data, do not have an absolute error calculated for each individual data
point.
When the two differently calibrated data sets have been merged together, 
the absolute errors 
on the NEWSIPS data can be used to estimate the error on the co-added
data.  

Grids of models were determined for each star by stepping through values
of log g, and finding the best    
fit temperature and normalisation [($R_\star/d$)$^2$] at each point. The
radius and mass were calculated using the
evolutionary models of Wood (1995), and the radii are then used to
estimate the distance from the normalisation parameter. The distances to
the primaries in each case are estimated from their spectral type and V
magnitude, and given in Table 3. The V magnitude of each white dwarf is
estimated from the model flux at 5500{\AA}. 
These results are all given in Table 5.

\subsection{White Dwarf Binaries - {\RO} data}

Once the temperature and gravity of each star has been determined, 
the {\RO} EUV and soft X-ray fluxes can give an indication of the level of
photospheric opacity in the white dwarf, by comparing them with predicted
values for a pure H atmosphere (e.g. Barstow et al. 1993). 
The {\RO} data is fitted independently from the {\IUE} data since
contamination from elements heavier than H and He only has a significant effect
at EUV and soft X-ray wavelengths.  
We fit the data from the two WFC filters, and the integrated count rate 
in the 0.1-0.28keV PSPC band, within which all the 
white dwarf soft X-ray flux is expected to lie. It is possible that
some EUV and X-ray emission might originate from the main sequence star
in these binaries. We therefore note the PSPC count rates at energies above
0.4keV as an indication of an active companion (Table 1).

We have fitted a set of fully line blanketed homogeneous H$+$He models,
computed by Koester (1991), to the {\RO} data, again using the XSPEC
spectral fitting programme. The model assumes
a homogeneous distribution of H$+$He, under LTE conditions, in
the range $-$8$>$log He/H$>$$-$3. A number of variables can
determine the predicted EUV/X-ray fluxes in the model - T, log g,
[($R_\star/d$)$^2$], H layer mass and HI, HeI and HeII columns. A valid
chi-squared analysis requires the number of degrees of freedom, $\nu$
(number of data points minus number of free parameters), to be greater
than or equal to one. Since we
fit only three independent data points we must use other
information to specify some of the parameters. Therefore, we 
use the T, log g and [($R_\star/d$)$^2$],
determined from the fit to the UV data and freeze these three
parameters during the fit. This time, however, the He/H ratio is
allowed to vary. We also make the reasonable
assumption that the local ISM is not highly ionised (therefore, there is
negligible HeII absorption) and that the HeI/H ratio is cosmic. Thus the
HI column can be estimated.

In using this technique the $\chi$$^2$ minimum  is often ill-defined.
We consider a good fit to the data to correspond to the probability that
a particular value of the 
reduced $\chi$$^2$ ($\chi_r$$^2$$=$$\chi$$^2$$/$$\nu$) can occur by chance
 to be 0.1 or greater (i.e. 90\%
confidence), and a bad fit 0.01 or less (99\%
confidence). The fits in between may not be very good, but cannot be
ruled out with high confidence. For $\nu$$=$1, as in this analysis,
then a good fit requires $\chi_r$$^2$ to be less than 2.71, but until the
value of $\chi_r$$^2$ exceeds 6.63 a model cannot be excluded with any
certainty. We therefore note all model fits to the {\RO} data for which
$\chi_r$$^2$ is less than 6.63.
Table 6 gives the HI column densities and He/H
ratios for the homogeneous fits to the {\RO} data. 

\subsection {Non-detections and Active Stars}

Emission features (e.g. CIV 1549{\AA} and CII 1335{\AA}) were detected in
the {\IUE} SWP spectra of four of the six targets where no white dwarf was seen.
A search through the recent literature showed
that other observers had also detected evidence for activity in these
stars at optical wavelengths. We analyse the UV, EUV and soft X-ray 
data for these stars.

The line fluxes of the emission features were measured using a 
simple Gaussian profile, fitted to each
line, after the continuum has first been subtracted (the continuum flux
is represented by a low degree polynomial). The measured line fluxes have
been compared with those of two known active stars, 
HD39587($\chi$$^1$ Ori, G0V) and HD126660($\theta$ Boo, F7V), 
also detected by the {\RO} WFC. Low resolution SWP spectra for these
stars exist in the {\IUE} archive, and  the line fluxes (Table 8) 
have been published by Ayres et al. (1995).  

Estimates of the ratio L$_{EUV}$/L$_{bol}$ have also been made as an
indicator of the level of activity. We have followed the method outlined
by Jeffries (1995). The S2 count rate is converted into a flux (f$_{EUV}$) 
in the 0.05-0.2 keV band using a uniform conversion rate of 1.5$\times$10$^{-10}$ erg
cm$^{-2}$ count$^{-1}$, assuming an effective coronal temperature of 
(5-10)$\times$10$^6$K. The bolometric luminosity $\it{m}$$_{bol}$ 
of each star is determined using the bolometric corrections given in
Allen (1973). L$_{EUV}$/L$_{bol}$ can then be calculated using
equation (1) of Jeffries (1995). These ratios are valid assuming neutral
hydrogen column densities less than $\sim$10$^{19}$cm$^{-2}$. As all four
active stars lie within a few tens of parsecs of the Sun, this is a
reasonable assumption.

The four active stars (CD$-$44 1025, HR1249, HR2468 and BD$-$00$^\circ$1462) 
were also detected 
by the PSPC in both the soft (S, 0.1-0.4keV) and hard (H, 0.4-2.4keV)
bands. We have therefore also calculated
L$_x$ and L$_x$/L$_{bol}$ from the total PSPC count rate, using the
hardness ratio 
(H$-$S/H$+$S) and the flux conversion factor given by Fleming et al. (1995).
We have also calculated L$_x$/L$_{bol}$ for the companion star in the WD binary
HD18131, which was detected in the PSPC hard band where no flux is
expected from the WD. A flux conversion factor for the hard band alone is
also given by Fleming et al. (1995).

The L$_{EUV}$/L$_{bol}$ and L$_x$/L$_{bol}$ ratios are presented in Table
9, along with estimates of L$_{EUV}$ and L$_x$.

\section {Discussion}

\subsection {White Dwarf Binaries}

\subsubsection {HD2133}

HD2133 (Figure 1) is listed as a V$=$9.7 F7V star in the SIMBAD database. If
these
parameters are correct, it lies at a distance of $\approx$ 160 parsecs. 
If the white dwarf is associated with HD2133, then 
the best fit T and log g from the UV data 
are 26,420$\pm$500K and 7.50 respectively (see Table 5). However, a fit to 
the {\RO} data points with these parameters yields a negligible column. In 
fact in this direction (l$=$305$^\circ$, b$=$-43$^\circ$) 
and for a distance of $\approx$ 160 parsecs, a HI column 
density of $\sim$10$^{19}$cm$^{-2}$ should be expected (Diamond, Jewell
and Ponman, 1995).

In the absence of optical data, we can use an {\IUE} LWP spectrum (LWP31757) 
of the main sequence star to estimate its spectral type by comparing it
with standard F stars in the {\IUE} archives. Unfortunately, we could find no
match in the database. We therefore suggest that we cannot be certain of 
the accuracy of the quoted V magnitude for this star. HD2133 would appear
to be closest in flux level and spectral shape to an F8V. However, a good
match to an F8V standard can only be achieved if we assume V$\approx$9.2.  
This would place the star $\sim$115pc away.

At that distance, our best fit pure-H white dwarf model is log g$=$8.25, 
T$\approx$28,700K and M$=$0.79M$_\odot$. A corresponding fit to the {\RO}
data (Table 6) gives an essentially pure H atmospheric composition, but
there is still a negligible hydrogen column density. Clearly, 
accurate optical spectroscopy and photometry is needed to better 
constrain the parameters of this system. 

\subsubsection {HD18131}

Vennes et al. (1995, hereafter V95) first reported the discovery of this 
binary, classifying HD18131 as K0IV, and concluded 
that together with the white dwarf it forms a physical pair at a
distance of 70$-$90pc. Based on the {\IUE} and {\euve} data sets, V95 
give the white
dwarf parameters as T$\approx$30,000K and log g$\approx$7.5, and the
interstellar column density as $\sim$10$^{19}$cm$^{-2}$. However, 
the PSPC count rates were not available to V95, so we 
include them here (Table 1).

We have adopted the same technique for analysing the {\IUE} and {\RO} data
as in the other binaries in this paper. We have no optical data for
HD18131 and so we use V95's spectral classification. 
At a distance of 70pc we find the
white dwarf parameters are T$=$31,130K, log g$=$8.0 and mass$=$0.65M$_\odot$
(see Table 5),
giving an age for the white dwarf of $\sim$10 million years. 
If the pair were further away at a distance of $\sim$90pc, log g$=$7.5,
T$=$29,290K
and mass$=$0.43M$_\odot$, and the white dwarf is slightly younger at 7
million years. However, for log g $\ge$8.0 the temperatures given in 
Table 5 are considerably lower than those measured by V95 
(e.g. for log g$=$9.0 we find T$\approx$35,000K but V95
give T$\approx$44,000K). 
These differences merely reflect the problems in deriving atmospheric
parameters for white dwarfs from IUE data alone. 
The Lyman $\alpha$ absorption
line in the SWP spectrum of HD18131 is partly filled-in by geocornal
emission (see Figure 2) and, as discussed below in Section 5.1.4, this can 
influence the values of the parameters measured by fitting this line. In
addition, a careful comparison at the short wavelength end of the
spectrum between V95's calibration  and our own  shows that
there is a small difference of $\sim$10\% in the flux level at 1300{\AA}.  
This may be attributable to the different 
method used by V95 to correct for the temporal degradation in detector 
sensitivity. V95 utilised an archival spectrum of the hot DA G191$-$B2B,
a method that involves assuming appropriate atmospheric 
parameters for that star, 
whereas we have used the correction of Bohlin and Grillmair (1988).

We include only the PSPC lower band
count rate (262$\pm$36 counts/ks) in our analysis of the {\RO} data set, 
assuming initially that only the white dwarf is responsible for this
flux. We find we can get good fits to the {\RO} data for log g$=$7.5 and 8.0
(see Table 6). 
At log g$=$8.0, the He/H ratio is $\approx$4$\times$10$^{-5}$ (i.e. virtually 
pure hydrogen) and the interstellar hydrogen column is
7.2$\times$10$^{18}$cm$^{-2}$, in agreement with V95.  
This is the only binary in the sample in this paper that has a detection
in the 0.4$-$2.4keV hard band of the PSPC (52$\pm$16 counts/ks). Since no
flux from the white dwarf is expected in this band, this soft X-ray flux
must be coming from the evolved K0 companion, confirming the conclusion
of V95 that the star is active. Although 
V95 found no CaII H \& K emission cores in their optical
spectrum of HD18131, the {\IUE} LWP spectrum reveals Mg II emission.
We derive, from the hard   
band count rate alone, an L$_{x}$$/$L$_{bol}$ ratio of 1.4$\times$10$^{-5}$, and
an X-ray luminosity L$_x$$\approx$3$\times$10$^{29}$erg s$^{-1}$ (assuming d$=$70pc).  

Could there be a contribution from the late type star to the PSPC soft
band flux? For example the white dwarf in the binary system   
HD217411 (Barstow et al. 1994) has a roughly similar
best fit temperature ($\approx$35,600K) and gravity (log g$=$8.2) to HD18131.
The hard band count
rate for this source is similar (56$\pm$15 counts/ks), but there is
a higher soft band rate (442$\pm$38
counts/ks). The authors found they could not fit the {\RO} data with
the parameters determined from a fit to the {\IUE} data. 
They conclude that the  G5 companion in this sytem is probably coronally
active and contaminating the soft PSPC band. 

In HD18131 we find 
the soft band flux can be fitted well without needing to account for a
contribution from the primary. 
However, if we make the reasonable assumption that there is as
much flux from the K0 subgiant below the carbon edge as there is above,
then by subtracting this contribution ($\sim$50 counts/ksec) 
we find we can only fit the {\RO} data points for log g$>$8.0. 
This would set an upper limit distance to the system of $\sim$70
parsecs. 

\subsubsection {RE J0357$+$283}

Jeffries, Burleigh and Robb (1996, hereafter JBR96) concluded that the
rapidly rotating  K2 dwarf star located at the centre of the
{\RO} source error box would have to be the most active star in the
galaxy to account for all the EUV and soft X-ray flux detected by the WFC.   
Instead, the authors predicted that a hidden white dwarf was
responsible for the EUV emission and 
they estimated, from the faint excess short wavelength flux in an 
{\IUE} LWP spectrum, that any degenerate companion must have a temperature 
between 30,000$-$40,000K, log g$=$7.5$-$8.0, and the H column density  
N$_H$$=$(2-6)$\times$10$^{19}$, assuming the distance to the system to be $\>$107pc. 

An {\IUE} SWP spectrum, obtained in August 1995, confirms this
identification (Figure 3). If the white dwarf is associated with the K
dwarf, then a fit to the UV data for
log g$=$7.9 gives approximately the minimum distance allowed. Fits at higher
gravities give distances inconsistent with that of the late-type star. At 
log g$=$7.9,  the temperature of the white dwarf T$=$30,960K, and 
mass$=$0.6M$_\odot$.
Fitting the {\RO} data with these parameters gives 
N$_H$$=$2.1$\times$10$^{19}$cm$^{-2}$, and the He/H ratio 
$\approx$10$^{-8}$. The atmosphere of this star can then be 
considered to be effectively pure H. We cannot fit the {\RO} data for log
g$<$7.9. This source is not detected in the PSPC hard band, and JBR96
argue that the vast majority of the soft X-ray counts must be coming from
the white dwarf and cannot be due to the K2V star, which shows no
evidence of chromospheric Mg II emission in the {\IUE} LWP spectrum. 

This is the second white dwarf/rapidly
rotating cool-star wide binary to be discovered. The other,
BD$+$08$^\circ$102 (RE J0044$+$093, Barstow et al. 1994, Kellett et al.
1995), consists of a K1$-$3V star with a spin period of $\sim$10 hrs,
and a white dwarf T$\approx$28,700K, log g$=$8.4 and mass$\approx$0.91M$_\odot$
(distance to the system is about 55pc). The K2V primary in RE J0357$+$283
has an even faster rotation period of 8.76 hrs. 
Both JBR96 and Kellett et al. (1995) find that the binary periods of these two
systems are likely to be measured in months or years, so it is unlikely
that there was ever common envelope evolution to spin-up the cool stars,
as is thought to have happened in the 12.5hr `pre-CV' K2V/hot white dwarf 
binary V471 Tau (Nelson \& Young 1970). Jeffries and Stevens (1996)
provide an alternative model to explain the spin-up in long-period 
binaries. They show that a significant amount of material and angular
momentum can be accreted from the slow, massive wind of an AGB star in a
detatched system. In this model final binary separations of up to $\sim$100AU 
are allowed. JBR96 also note that the cool star has a slightly enlarged
radius. As the white dwarf is relatively young, $\sim$8 million years in
the log g$=$7.9 model, then this can be explained by the cool star having
yet to settle at a new main sequence radius after the accretion of a
large amount of mass.

\subsubsection {BD$+$27$^\circ$1888}

The white dwarf companion to BD$+$27$^\circ$1888 was discovered with {\IUE} 
(SWP49780) in January 1994. In our initial studies of the white dwarf 
we utilised this spectrum because SWP49779, 
obtained at the same time, was overexposed in
places by as much as a factor of 2. Unfortunately, the hydrogen Lyman
$\alpha$ absorption 
line in SWP49780 has been heavily filled in by geocoronal emission, and in 
addition there is a reseau mark (giving a zero reading across several 
channels) on one wing, making modelling extremely difficult, so we 
reobserved the star in the SWP and LWP cameras in November 1995.
These two spectra, SWP56261 (Figure 4) and LWP31785(Figure 12), 
were both extracted with the NEWSIPS 
calibration. In the SWP this largely removes the contamination from 
the geocoronal Lyman $\alpha$ line, and thus the hydrogen absorption line
can be 
fitted more accurately. In addition there are no zero points due to reseau 
marks, and the NEWSIPS calibration also provides an absolute 
error for each individual data point. Figure 14 compares the NEWSIPS
extracted spectrum with the earlier one, and clearly in the NEWSIPS
spectrum there are many more
uncontaminated data points to fit in the crucial  Lyman $\alpha$ region. 
Fitting the NEWSIPS data alone, we find 
that the temperature is actually higher for each value of log g by
$\sim$6000K at 
the lower end of the range and by $\sim$10,000K at the upper end. Table 7 
shows a comparison between the fitted temperatures for both spectra, 
stepping up through log g. 

The differences in the fits between the two spectra is most likely due to
the poor quality of the SWP49780 spectrum, in which it is necessary to
remove, before modelling, 
a large number of the most vital data points in the Lyman $\alpha$
region. Therefore, rather than attempt to co-add the spectra, 
we decided to use the NEWSIPS calibrated 
spectrum alone. The results in Table 5 are all derived from this data set.

There is some discrepency between the V magnitude given for the main
sequence star in the SIMBAD database (9.6) and 
the Guide Star Catalogue (GSC, 9.08). We
have made an estimate of V from our optical spectrum
(Figure 4), and agree with the GSC 
value of $\approx$ 9.1. Unfortunately, there was cloud during
our observation and therefore the absolute flux values of the spectrum
are not totally reliable. 
We adopt a spectral classification of A8V-F2V from this spectrum for the
primary, although SIMBAD lists it as a G5. 
However, it is clear from Figure 4 that there is considerable flux from
this star in the {\IUE} SWP spectrum, where a G5 would not be seen. 
We also attempted to match the LWP spectrum (LWP31785) to
stars of known spectral type in the {\IUE} database. From this method we
find the star is in fact closest in shape and flux to an A8V (Figure 12). 
From a quick glance through an atlas of optical spectra (e.g.
Jacoby, Hunter and Christian 1984) it is immediately apparant that the
differences in spectral shape between late A type stars and early Fs is
very subtle. However, Figure 12 shows that at UV wavelengths the
difference in flux level from one subclass to the next is substantial. In
Figure 12 we compare the LWP spectrum against examples of F2V 
and A8V spectra from the archive, each scaled for the differences in V
magnitude, and BD$+$27$^\circ$1888 most closely resembles an A8V.  

If the primary is in the range of spectral types A8-F2V and assuming V$=$9.1, 
it lies between 185-218 parsecs away. The closest model fit we have
for a white dwarf at about that distance is T$=$34,000K, log 
g$=$7.25, and M$=$0.39M$_\odot$. This mass estimate is surprisingly low,
given that if this is a true binary then the white dwarf must have
evolved from a progenitor more massive than an A8V. However, we must be
careful not to over-interpret these results, particularly given the
difficulties in fitting {\IUE} data unambiguously, as discussed above. 
Fitting the {\RO} data with these parameters, we find an essentially 
pure H atmosphere and a column of 3.3$\times$10$^{19}$cm$^{-2}$. The
entire spectrum of BD$+$27$^\circ$1888, from the far-UV to the optical,
is displayed in Figure 15. 

\subsubsection {RE J1027$+$322}

The discovery of this WD$+$MS binary was first reported by Genova et al. 
(1995, hereafter G95). 
The authors studied the entire field of the {\RO}/{\euve} source, and concluded 
that although it was more likely that the white dwarf was the sole source of
the EUV flux, there may be a contribution from a QSO in the field. G95 
used only a single spectrum, SWP49778, originally obtained by us in
January 1994. Subsequently, we have obtained two further exposures,
SWP49793 and SWP54501, and 
co-added all three to give the results we present here. This higher
quality data allows us to better constrain the white dwarf parameters. An 
intense emission feature is seen in SWP49778 at ~1700{\AA}, but there 
are no known strong emission features at this wavelength. G95 consider
that it may well be spurious, and it is not seen in either 
SWP49793, taken $\approx$
48hrs later, or in SWP54501. After examination of the photowrite image of
the detector plate, we conclude that this emission feature is due to a
cosmic ray hit.

G95 estimate the spectral type of the main sequence star to be
G2($\pm$2)V, and that it lies at a distance of 380$-$550pc. If the white
dwarf is indeed associated with this star, then we find, by fitting the
{\IUE} data, that it must have a surface gravity log g of 7.0-7.5 and a
temperature of around 30$-$33,000K. 
However, we also find from fitting the {\RO} data that a good fit can only be
achieved at log g$=$7.5 or higher, although fits at log g$=$7.0 and 7.25
cannot be completely excluded. At log g$=$7.5, T$=$32,440K, log g$=$7.5,  
M$=$0.44M$_\odot$, 
the column density N$_H$$=$1.05$\times$10$^{19}$cm$^{-2}$, and the atmosphere can
be assumed to be essentially pure H.  

\subsection {Non-detections}

\subsubsection {AG$+$68 14}

This V$=$10.3 star, listed in SIMBAD as an F8, was observed as the
potential counterpart to the 2RE EUV source 0014$+$691. It was selected as a
possible white dwarf binary on the basis that it is a very soft EUV
source. The UV spectrum (Figure 6, SWP52807) 
shows virtually no flux above the background, except possibly at the long 
wavelength end. Comparing with F8 stars in the {\IUE} spectral atlas (Wu et
al. 1992), we would 
expect to see little flux from the star in the SWP camera. We conclude
that there is no white dwarf companion to this star, and in the absence of  
any emission features in the far UV, and the non-detection of the source
by the PSPC, we also conclude that it is probably
not active. The field of this star needs re-examining in the optical to find
the true EUV source.

\subsubsection {CD-44 1025}

This star is a PSPC X-ray source with a detection in the 
hard band (575$\pm$41 cs$^{-1}$). Early references (e.g. Malaroda 1973)
claim it is a spectroscopic binary (F3V$+$A8V). The 2RE catalogue 
lists it as an F7III. Mason et al. (1995) observed the star during the WFC 
identification programme. They estimate the apparent EUV to optical flux
ratio 
$=$2.25, and the equivalent width of the Ca H 3933{\AA} chromospheric emission 
line $\approx$0.1{\AA}. The authors note that the emission cores are
variable in strength. They only class the star as an F type.

We observed the star in both the {\IUE} SWP and LWP cameras (Figure 7). Our 
two SWP spectra (SWP56046 and SWP56047) were taken consecutively with
15 minute exposures and co-added to improve the signal/noise. Emission lines of
CIV 1549{\AA} and HeII 1640{\AA} are visible; measurements of  
these lines are presented in Table 8, where we also compare them with 
similar lines in low resolution {\IUE} SWP spectra of HD126660 (F7V) and 
HD39587 (G0V), two known active stars in the 
WFC catalogue (RE J1425$+$515 and RE J0554$+$201). 

As with BD$+$27$^\circ$1888 (above) we attempted to match the LWP
spectrum (LWP31570) to stars of known spectral type in the {\IUE} archive.
Figure 13 shows that at these UV wavelengths the star is closest in flux
level and shape to an A8V. However, below $\sim$2800{\AA} the flux levels
are not quite matched and this could be an indication that this is indeed
a binary. In Table 9 we present
measurements of the EUV and X-ray luminosities, assuming the spectral
type of this star to be (a) F3V and (b) A8V. 

\subsubsection {HR1249}

Until its discovery as an EUV source in the WFC survey, this star 
(RE J0402$-$001) was not known to be 
active. However, it is detected in the 0.4$-$2.4keV band of the PSPC, and 
further evidence of activity was observed optically by Mason et al. (1995) 
during the WFC source identification programme. They   
estimated the EUV to optical flux ratio$=$2.6, and measured the equivalent 
width of the CaII H 3393{\AA} emission core$=$0.01{\AA}. 
However, they note that they observed this weak 
emission core on only one occasion. The star has also been studied by 
Jeffries and Jewell (1993) who classify it as F6V. 

Figure 8 shows SWP49792, and clearly there is no white dwarf present.
However, chromospheric emission lines of CIV 1549{\AA}, SiIV {1397\AA}, 
CII 1335{\AA} and HeII 1640{\AA} are visible.
Measurements of 
these lines are presented in Table 8, and measurements of the X-ray and
EUV luminosities are given in Table 9. Comparing with the known
active star WFC sources HD39587 and HD126660, we conclude that 
HR1249 is indeed the source of the EUV radiation.

\subsubsection {HR2468 (HD48189)}

This star (alternatively HD48189) 
was not well studied prior to the WFC survey, and not known to be
magnetically active.  Jeffries and
Jewell (1993) find it is a double star, and class the two
components as G0V and K3V respectively. SIMBAD lists a combined spectral type
of G1.5V. The EUV source (RE 
J0637$-$613) has proved to be a young lithium-rich object, and  
Jeffries (1995) measures the equivalent width of the LiI 6708{\AA} line $=$ 
133m{\AA}.

The {\IUE} SWP spectrum (Figure 9, SWP52802) shows no evidence for a white dwarf 
companion, but there are CIV 1550{\AA}, CII 1335{\AA} and HeII 1640{\AA} 
emission lines visible. 
Measurements of these lines are presented in Table 8. Measurements of the
EUV and X-ray luminosities in Table 9 are made assuming a G1.5V spectral
classification. 

\subsubsection {BD-00$^\circ$1462}

This star was observed three times as part of the WFC optical
identification 
programme of Mason et al. (1995), although the authors failed to find emission 
cores in either CaII H\&K or H$\alpha$, with an upper limit to the equivalent 
width of 0.05{\AA}. As an active F2V star, BD-00$^\circ$1462 has been well 
studied in the far UV and optical (e.g. Oranje \& Zwaan 1985, Simon \& 
Landsman 1991, Andersson \& Edvardsson 1994) and is very likely the
optical counterpart to RE J0650$-$003. 
In the far-UV, emission cores are seen in the 
MgII H \& K lines in high resolution LWP spectra. We obtained one low
resolution SWP spectrum, SWP52801, although earlier spectra exist in the archive.
Figure 10 shows the best of these, SWP8200, with an inset of SWP52801
showing a possible CIV 1549{\AA} emission feature. 
There is clearly no white dwarf. This star is also an X-ray source
(see Table 9). 

\subsubsection {HD166435}

Observations of this star in the optical have revealed little evidence
of activity, hence it was selected 
by us as a likely candidate for an unresolved white dwarf binary. The
star is 
classified in SIMBAD as G0. Observations by Mason et al. (1995) showed it
to be, at 
most, mildly active. They measured an equivalent width of the CaII H
3393{\AA} emission line of 0.07{\AA}, but detected no emission core in
H$\alpha$. 
Mullis \& Bopp (1994)  also concluded that there was insufficient evidence to 
prove that this 
star is chromospherically active. They found no emission core in H$\alpha$,
and only a small core in CaII 8542{\AA} which they were not convinced was real.

We observed HD166435 in low resolution with the {\IUE} SWP camera 
in August 1995 (Figure 11, SWP55658).
There is clearly no white dwarf companion, and no obvious emission features
although ,there is a hint of a CIV 1549{\AA} emission feature rising above the
background noise with a peak flux of $\sim$3$\times$10$^{-14}$ ergs cm$^{-2}$
sec$^{-1}$.

The WFC source is, however, coincident with a PSPC source including a
0.4$-$2.4keV hard band detection. A white dwarf can be excluded as the origin of
this hard band radiation, which suggests that HD166435 may indeed be coronally
active. If we assume the source is HD166435, then the X-ray
luminosity of this star L$_x$$=$9.3$\times$10$^{29}$, 
L$_x$/L$_{bol}$$=$1.5$\times$10$^{-4}$, and L$_{EUV}$/L$_{bol}$$=$1.6$\times$10$^{-4}$. 
Comparing with the other stars in Table 9, this would make HD166435 
a reasonably active object. 
We suggest that further high resolution optical spectroscopy of this
object is required in order to unambiguously determine whether it is
an active star, and the field of this source may also need re-examining.

\section {Summary}

We have discovered three more hot white dwarfs in unresolved
non-interacting binary systems with main sequence companions (HD2133, RE
J0357$+$283 and BD$+$27$^\circ$1888). In addition, we have independently
identified a fourth system, RE J1027$+$322, previously reported in the
literature by Genova et al. (1995). 
This brings the total number of such binaries found as a result of the  
EUV surveys to
fifteen, in addition to previously identified systems also
seen by the {\RO} WFC (e.g. V471 Tauri)  
\footnote{Since this paper was first submitted,
Burleigh and Barstow (1997) have discovered another
WD$+$MS binary, RE J0500$-$362. This system will be discussed in more
detail in a future paper.}.  
Including WD+dM pre-CV and wide pairs, there
are now $>$30 white dwarfs in non-interacting binaries that have been 
detected in the EUV. 
                                
Between 50 and 80\% of all stars are believed to lie in binary or
multiple systems. 
However, in the {\RO} X-ray catalogue of white dwarfs 
(Fleming et al. 1996), only 23\% of the 176 objects 
are detected in binary systems. In addition, 
75\% of the $\sim$120 {\RO} WFC white dwarfs 
appear to be single stars. It is reasonable 
to believe, then, that there are more white dwarf binaries in the 
EUV catalogues waiting to be discovered.  

However, in our continuing {\IUE}
programme we are now observing much fainter EUV sources and less obvious
candidates, and identifying new systems is becoming increasingly
difficult. There are still, though, some excellent candidates to be
targeted. From their WFC count rates and S2/S1 ratios it is likely that 
two bright EUV sources which appear to be 
associated with inactive B stars (HR3665 and HR2875) will have hot white
dwarf companions. These white dwarfs, should they exist, will have to be 
identified 
spectroscopically with {\euve} since the B stars will dominate the {\IUE}
wavelength ranges.

\section*{Acknowledgements} MRB and MAB acknowledge the support of PPARC, UK. 
TAF acknowledges support from NASA under grant NAGW-3160. 
We wish to thank Detlev Koester for the use of his white
dwarf model atmosphere grids, and Jay Holberg and Jim Collins at the
University of Arizona with their help in obtaining and reducing some of
the data. MRB wishes to thank the staff at {\IUE} Vilspa for their help
and co-operation during his visits there, in particular John Fernley,
Constance la Dous and Richard Monier. This research has made use of the
Simbad database operated at CDS, Strasbourg, France.

\newpage

\section*{Figure Captions}

\subsection*{White Dwarf Binaries}

Figure 1. Co-added low resoltuion {\IUE} SWP spectrum of HD2133
(SWP55659$+$SWP56231), compared to a predicted pure H white dwarf spectrum 
for log g$=$8.25
and T$=$28,700K. Note that there may be  a small contribution to the flux at
the long wavelength end from the F7V$-$F8V companion. 

Figure 2. Low resolution {\IUE} SWP spectrum of HD18131 (SWP52158),
compared to a predicted pure H white dwarf spectrum for log g$=$8.0 and
T$=$31,130K.

Figure 3. Low resolution {\IUE} SWP spectrum of RE J0357$+$283
(SWP55660), compared to a predicted pure H white dwarf spectrum for log
g$=$7.9 and T$=$30,960K.

Figure 4. Low resolution {\IUE} SWP spectrum of BD$+$27$^\circ$1888
(SWP56261), compared to a predicted pure H white dwarf spectrum for log
g$=$7.25 and T$=$34,130K. Note the contribution to the flux longwards of
$\sim$1600{\AA} from the A8V$-$F2V companion.

Figure 5. Co-added low resolution {\IUE} SWP spectrum of RE J1027$+$322
(SWP49778$+$SWP49730$+$SWP54501), compared to a predicted pure H white
dwarf spectrum for log g$=$7.5 and T$=$32,440K. 

\subsection*{Non-detections}

Figure 6. Low resolution {\IUE} SWP spectrum of AG+68 14 (SWP52807).

Figure 7. Low resolution {\IUE} spectrum of CD-44 1025 (A8V$-$F3V, coadded
SWP56046$+$56047 and LWP31570). 
Inset, the co-added SWP spectrum showing more clearly the CIV 1549{\AA} 
emission feature.

Figure 8. Low resolution {\IUE} SWP spectrum of HR1249 (F6V, SWP49792).
Emission lines of CII 1335{\AA}, SiIV 1397{\AA}, CIV 1549{\AA} and
possibly HeII 1640{\AA} are visible. 

Figure 9. Low resolution {\IUE} SWP spectrum of HR2468 (G1.5V,
SWP52802).Emission lines of CII 1335{\AA}, CIV 1549{\AA} and
possibly HeII 1640{\AA} are visible.

Figure 10. Low resolution {\IUE} SWP spectrum of BD-00$^\circ$ 1462 (F2V, 
SWP8210). Inset, part of SWP52801, showing a possible CIV 
1549{\AA} emission feature.  

Figure 11. Low resolution {\IUE} SWP spectrum of HD166435 (G0, SWP55658).

\subsection*{Other diagrams}

Figure 12. Low resolution 
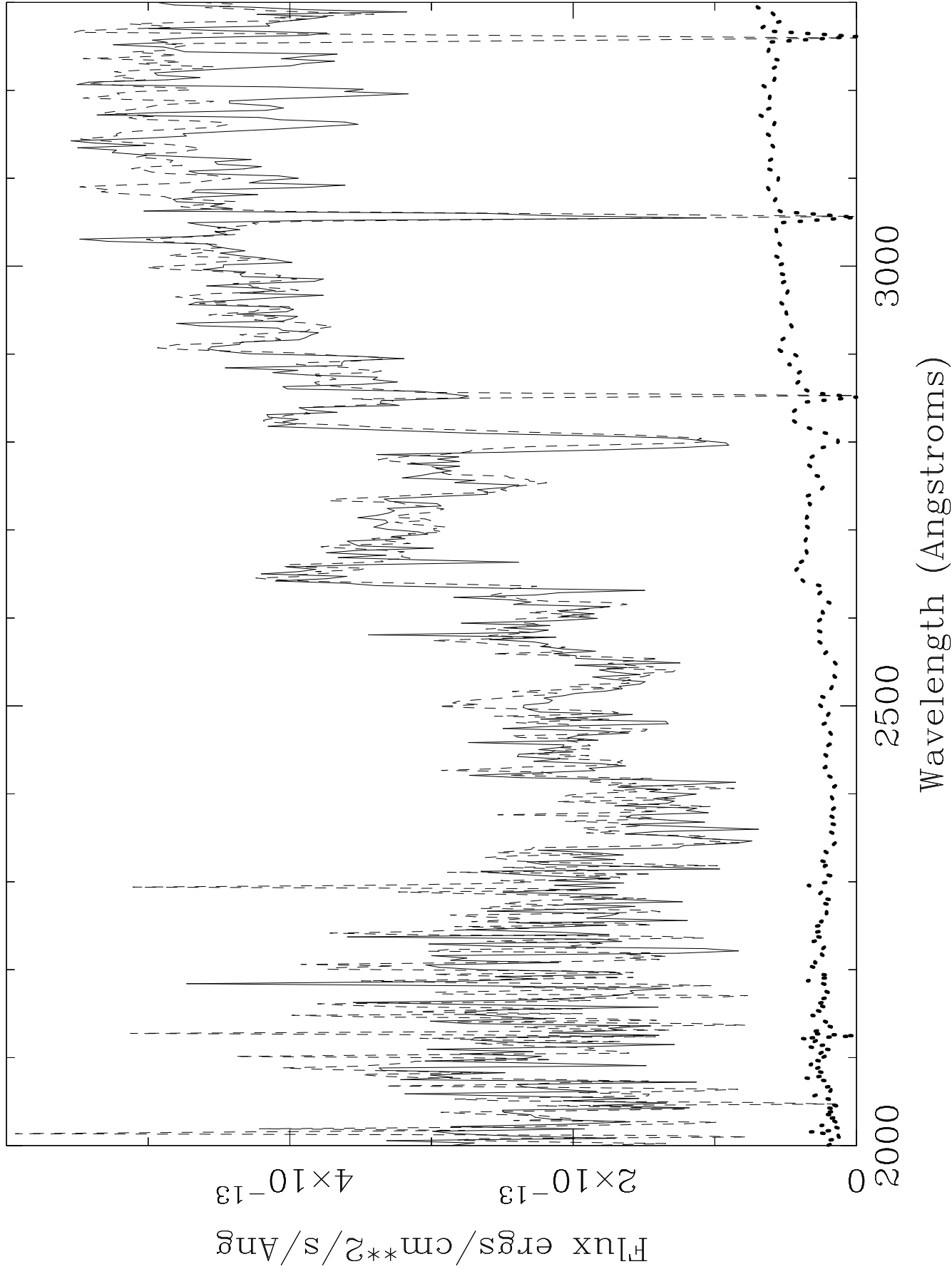
{\IUE} LWP spectrum of BD+27$^\circ$ 1888 (LWP31785, solid line)
and comparison spectra - HD28910 (LWP27455, A8V, dashed line) and HD29875
(LWP20865, F2V, dotted line), scaled for differences in magnitude
(assuming V$=$9.1 for BD+27$^\circ$ 1888).

Figure 13. {\IUE} LWP spectrum of CD-44 1025 (LWP31570, solid line)
and comparison spectrum - HD28910 (LWP27455, A8V, dashed line),
scaled for difference in magnitude.

Figure 14. Comparison of NEWSIPS SWP spectrum of BD+27$^\circ$ 1888
(SWP56261, upper) with the older IUESIPS-extracted SWP49780 (lower). The flux
levels are arbitrary. Note the presence of a strong geocoronal Lyman
$\alpha$ emission line in SWP49780. 

Figure 15. {\IUE} SWP and LWP spectra of BD$+$27$^\circ$1888 (WD$+$F2V), 
displayed together with an optical spectrum and a pure H model white dwarf
spectrum for T$=$34,130K and log g$=$7.25.
The white dwarf can be seen
emerging from the glare of its companion shortwards of $\sim$1600{\AA}. 
This diagram clearly illustrates 
that the white dwarfs in these binaries are undetectable at optical 
wavelengths and can only be seen in the far-UV.
 
\end{document}